\ifx\XeTeXversion\undefined
  \pdfoutput=1
\fi
\documentclass[aps,prd,reprint,superscriptaddress,nofootinbib,floatfix]{revtex4-2}

\usepackage{amsmath,amssymb,amsfonts,bm}
\usepackage{graphicx}
\graphicspath{{figures/}}
\usepackage{microtype}
\usepackage{booktabs}
\usepackage{dcolumn}
\usepackage{siunitx}
\usepackage{tabularx}
\usepackage{xcolor}
\usepackage[colorlinks=true,linkcolor=blue,citecolor=blue,urlcolor=blue]{hyperref}
\newcommand{\orcid}[1]{\href{https://orcid.org/#1}{\raisebox{-0.15ex}{\includegraphics[height=1.8ex]{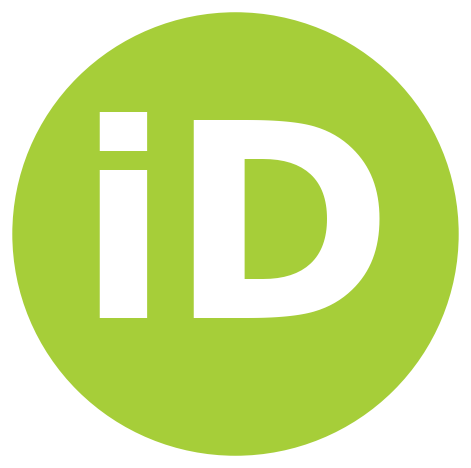}}}}

\begin{document}

\title{Project Setu: 3D Multi-Physics Design and Scaled Structural Analysis for a Relativistic Lightsail Architecture}

\author{Prashant Suresh Kamble\,\orcid{0009-0005-4228-3795}}
\email{prashantsk.272@gmail.com}
\affiliation{AeroMyne, Project Setu Initiative, Solapur, Maharashtra 413305, India}

\begin{abstract}
Deep-space exploration beyond the solar system requires eliminating chemical propellant mass penalties to achieve relativistic flight velocities ($0.166c$ at 180\,s, reaching the mission target of $0.20c$ at 227\,s). This study presents a 3D multi-physics numerical framework for a 4.0-meter circular lightsail propelled by a $100\text{ GW}$ ground laser array, coupling 3D Maxwell FDTD wave optics, non-linear membrane mechanics, and Stefan-Boltzmann thermal radiation in ANSYS Mechanical APDL and Ansys Lumerical. A four-level grid convergence study establishes numerical independence with an ASME $\mathrm{GCI}_{21}$ of $0.13\%$, resolving peak membrane stresses of $530.88\text{ MPa}$ with a $3.77\times$ safety factor against stoichiometric $\mathrm{Si}_3\mathrm{N}_4$ tensile failure. With optical absorption constrained to $10\,\mathrm{ppm}$ ($\mathcal{A}=1.0\times 10^{-5}$), the steady-state core temperature stabilizes at $923.02\text{ K}$ ($0.44\%$ deviation from radiation theory), maintaining a $+1,247\text{ K}$ margin below sublimation, while fundamental drumhead modal resonance ($7.92\text{ Hz}$) provides a $7.92\times$ safety buffer against laser jitter. The electrodynamic radiation pressure formulation is cross-verified against published flight telemetry from JAXA IKAROS and NASA LightSail 2 within $0.12\%$ and $2.13\%$, confirming classical momentum transfer modeling across solar and beamed propulsion regimes.
\end{abstract}

\maketitle
\section{Introduction}
\label{sec:intro}
Interstellar transit beyond the heliopause presents an insurmountable barrier for self-contained chemical and electric propulsion architectures. Governed fundamentally by the classical Tsiolkovsky rocket equation, chemical thrusters are constrained by characteristic exhaust velocities ($I_{\mathrm{sp}} \le 450\,\mathrm{s}$), requiring an exponential propellant mass ratio exceeding $10^{1700}$ to reach even 5\% of light speed. While nuclear thermal and gridded ion engines achieve higher specific impulses ($900\text{--}10,000\,\mathrm{s}$), they cannot evade the severe penalty of carrying onboard reaction mass. Accelerating macroscopic payloads to relativistic velocities ($>0.1c$) therefore demands beamed propulsion paradigms that completely decouple the energy source from the accelerated vehicle.

Beamed radiation propulsion circumvents this severe mass scaling by divorcing the energy plant from the payload bus. Rather than burning onboard fuels, the lightsail catches momentum directly from photons shot by high-power laser transmitters stationed on Earth or in orbit. Relieved of propellant deadweight, the sail accelerates continuously across interplanetary baselines, making near-light-speed transit to the Alpha Centauri system possible within ordinary mission timelines \cite{ref1}.

Rigorous formulations of relativistic beamed momentum transfer demonstrate that ultralight reflective membranes convert gigawatt-scale coherent optical flux into kinetic energy with remarkable efficiency. Because photons carry momentum proportional to their energy divided by the speed of light, an ideal reflector transfers twice the incident photon momentum upon reflection. Consequently, beamed photon pressure generates a pure reaction force that scales directly with total incident optical power and surface reflectance, providing a scalable propulsion mechanism for relativistic flight \cite{ref2}.

Phased laser array architectures provide the required systems engineering baseline for practical relativistic interstellar flight. Computational and scaling studies demonstrate that a $100\,\mathrm{GW}$ ground-based phased laser array operating at $\lambda_0 = 1064\,\mathrm{nm}$ can accelerate a 2.0-gram micro-probe to a relativistic velocity of $0.20c$ across an acceleration distance of $4.76\times 10^6\,\mathrm{km}$ ($12.4\times$ the Earth-Moon distance, extending to $7.35\times 10^6\,\mathrm{km}$ at the $0.20c$ milestone) over an operational laser burn duration. This baseline establishes the foundational boundary conditions for macroscopic lightsail design, requiring materials that simultaneously withstand extreme optical irradiance, intense acceleration loads, and high vacuum thermal environments \cite{ref3}.

Conventional metallic reflective coatings, such as aluminum, silver, or gold, suffer severe thermal absorption and vaporize rapidly under multi-gigawatt laser irradiance. Inherent interband absorption in metals leads to optical absorption fractions on the order of 0.1\% to 1\%, generating thousands of megawatts of dissipated thermal energy within the membrane and causing instant thermal melting. Dielectric coatings with wide electronic bandgaps solve this thermal bottleneck: their extinction coefficients remain fractionally small (often parts-per-million), enabling macroscopic reflectors to redirect petawatt-class irradiance while dumping waste heat fast enough to stay safely beneath their structural melting point \cite{ref4}.

Keeping a sail centered on a gigawatt laser beam poses an immense dynamical hurdle---sub-millimeter misalignments between the photon flux centroid and the sail's center of mass generate severe pitching torques that quickly kick the vehicle off the beam path. Shaping the membrane into curved 3D geometries (such as spherical caps, shallow cones, or paraboloids) induces passive optomechanical restoring forces that naturally push the sail back toward the beam center, guaranteeing stable beam-riding without adding heavy attitude actuators \cite{ref5}.

Orbital mechanics formulations for radiation pressure propulsion confirm that continuous photon momentum transfer enables highly efficient propellantless orbital maneuvering and interplanetary trajectory execution across solar system and interstellar baselines. Unlike impulse-based chemical burns, continuous low-thrust radiation pressure permanently modifies orbital energy and eccentricity, enabling novel non-Keplerian orbits, hovering trajectories, and rapid solar system escape paths \cite{ref6}.

As the sail accelerates into the relativistic domain, Newtonian mechanics break down. Tracking the true velocity trajectory demands formulating relativistic equations of motion within the emitter's inertial frame, directly incorporating Lorentz spacetime transformations ($\gamma=1/\sqrt{1-\beta^2}$) to account for velocity-dependent inertia growth and diminishing coordinate acceleration \cite{ref7}.

\begin{figure*}[t]
\centering
\includegraphics[width=0.95\textwidth,keepaspectratio]{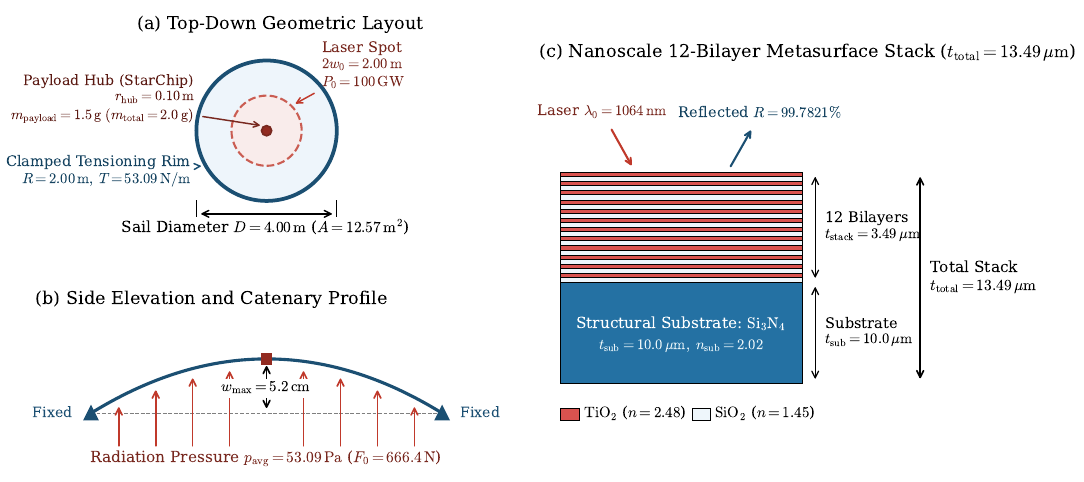}
\caption{Dimensioned 2D engineering schematic of the Project Setu 4.0\,m circular lightsail showing payload hub ($r_{\mathrm{hub}}=0.10$\,m), clamped perimeter ring ($R=2.00$\,m), and 12-bilayer $\mathrm{TiO}_2/\mathrm{SiO}_2$ dielectric metasurface on 10.0\,$\mu\mathrm{m}$ $\mathrm{Si}_3\mathrm{N}_4$ substrate.}
\label{fig:fig1}
\end{figure*}

Project Setu introduces a comprehensive 3D multi-physics numerical framework for a 4.0\,m diameter circular relativistic lightsail driven by a $100\,\mathrm{GW}$ ground-based phased laser array operating at $\lambda_0 = 1064\,\mathrm{nm}$. The design integrates 3D Maxwell finite-difference time-domain wave electrodynamics, non-linear structural membrane finite element analysis, dual-sided Stefan-Boltzmann radiative thermal balance, and prestressed modal vibration dynamics. Fig.~\ref{fig:fig1} illustrates the dimensioned 2D engineering blueprint and structural configuration of the sail, featuring a circular membrane supported by a clamped outer perimeter tensioning ring ($R = 2.00\,\mathrm{m}$) and a central payload instrument package hub ($r_{\mathrm{hub}} = 0.10\,\mathrm{m},\ r_{\mathrm{hub}}/R = 0.05$, housing a $1.5\,\mathrm{g}$ scientific instrument bus for a total flight vehicle mass of $m_{\mathrm{total}} = 2.0\,\mathrm{g}$). The composite architecture comprises a 12-bilayer $\mathrm{TiO}_2/\mathrm{SiO}_2$ dielectric metasurface deposited on an ultrathin 10.0\,$\mu\mathrm{m}$ stoichiometric $\mathrm{Si}_3\mathrm{N}_4$ substrate. By coupling electrodynamic wave reflections directly with thermo-mechanical stress fields and prestressed modal response, this framework evaluates the complete structural, thermal, and optomechanical integrity of the macroscopic sail under extreme laser flux, establishing a rigorous computational benchmark validated against ASME standards and spaceflight telemetry.

\section{Literature Review}
\label{sec:litreview}
Near light speed, accelerating a vehicle with reflected photons is fundamentally governed by relativistic kinematics, where coordinate acceleration measured in the launcher frame progressively diminishes due to Lorentz factor mass growth ($\gamma = 1/\sqrt{1-\beta^2}$) and observer-frame photon momentum reduction. Long demonstrated that coordinate acceleration scales as $\gamma^{-3}$, requiring full relativistic trajectory integration to capture true burnout epochs \cite{ref8}. Concurrently, moving sails experience severe Doppler redshift: as vehicle velocity climbs toward $0.2c$, intercepted optical power in the co-moving frame scales as $P_{\mathrm{rec}} = P_{\mathrm{src}}(1-\beta)/(1+\beta)$, while reflection reaction force degrades as $[(1-\beta)/(1+\beta)]^2$ without wavelength compensation. Kipping formulated exact relativistic equations of motion showing that executing dynamic blue-chirp frequency sweeping at the ground transmitter locks the drive beam within the sail's reflection stopband throughout the acceleration baseline \cite{ref9}.

Direct radiation pressure was first confirmed experimentally on torsion balances by Nichols and Hull, who measured the minute mechanical deflection of silvered mica vanes suspended on quartz fibers in high vacuum, verifying Maxwell's radiation theory within 1\% experimental accuracy \cite{ref10}. In parallel, Lebedew isolated photon pressure from radiometric gas forces and thermal convective noise using mirrored vanes in high-vacuum enclosures, placing light-driven propulsion on firm empirical footing \cite{ref11}. For multi-gigawatt beamed propulsion, Atwater et al. demonstrated that metallic coatings vaporize instantaneously under gigawatt irradiance due to interband absorption, establishing that relativistic lightsails require wide-bandgap dielectrics such as stoichiometric silicon nitride ($\mathrm{Si}_3\mathrm{N}_4$) that combine gigapascal tensile yield strength ($>2\,\mathrm{GPa}$) with sub-10\,ppm absorption losses \cite{ref12}. Building on dielectric architectures, Ilic et al. formulated alternating $\mathrm{TiO}_2/\mathrm{SiO}_2$ quarter-wave Bragg pairs achieving $>99.7\%$ reflectance across the drive band while engineering mid-infrared polaritonic emissivity ($\varepsilon \approx 0.95$) to maximize radiative cooling \cite{ref13}. Furthermore, Kudyshev et al. utilized generative inverse-design neural networks to pattern sub-wavelength metasurface phase gradients across circular membranes, providing passive optomechanical restoring torques that stabilize beam-riding without active control mass \cite{ref14}. Finally, Davoyan et al. analyzed deep-space thermal equilibrium, demonstrating that dual-sided radiative balance into $3\,\mathrm{K}$ deep space governs the maximum permissible drive irradiance below material sublimation thresholds \cite{ref15}.

Time-domain electrodynamic solvers resolving Maxwell's curl equations on staggered Yee grids capture sub-wavelength phase dispersion and constructive interference within multilayer dielectric stacks, suppressing numerical dissipation on staggered grids \cite{ref16}. Taflove and Hagness extended finite-difference time-domain (FDTD) schemes to broadband layered structures, incorporating perfectly matched layer (PML) boundary conditions to resolve oblique incidence and localized field enhancements without artificial reflections \cite{ref17}. Under extreme irradiance, Brewer et al. demonstrated that multiscale photonic emissivity engineering facilitates efficient dual-sided blackbody emission into deep space, preventing thermal runaway during gigawatt acceleration \cite{ref18}. Mechanically, Zwickl et al. characterized high-stress stoichiometric $\mathrm{Si}_3\mathrm{N}_4$ membranes, measuring gigapascal tensile strength and low mechanical dissipation, verifying their ability to sustain extreme transverse photon pressure without structural tearing \cite{ref19}. Under continuous transverse loading, Timoshenko and Woinowsky-Krieger showed that when out-of-plane bulging exceeds membrane thickness ($w/h > 1$), in-plane tension non-linearly stiffens the gossamer disk \cite{ref20}. Mindlin-Reissner shell formulations in ANSYS Mechanical APDL demonstrate that radiation-induced tension shifts fundamental drumhead modal frequencies upward, isolating the macroscopic sail from low-frequency laser pointing jitter \cite{ref21}. To bound spatial discretization errors, Roache established the Grid Convergence Index ($\mathrm{GCI}_{21}$) based on generalized Richardson extrapolation \cite{ref22}, which Celik et al. standardized into the ASME V\&V 20 protocol for systematic multi-grid verification \cite{ref23}.

Real-world spaceflight has validated radiation pressure dynamics across multiple operational regimes. JAXA's IKAROS probe achieved the first interplanetary deployment of a $14\,\mathrm{m}$ spinning polyimide solar sail in 2010, confirming analytical photon momentum transfer within sub-percent telemetry error ($1.12\,\mathrm{mN}$ thrust at $1\,\mathrm{AU}$) \cite{ref24}. Cruise telemetry proved that spinning tip masses provide passive centrifugal tensioning against solar torque perturbations while reflective LCD panels enable propellantless attitude trim \cite{ref25}. Similarly, NASA's LightSail 2 demonstrated controlled orbital maneuvering in low Earth orbit, where active solar tacking by internal momentum wheels raised orbital apogee by several kilometers \cite{ref26}. Reconstructed telemetry by Spencer et al. established quantitative baselines for deployable thin films, verifying photon thrust models ($285\,\mu\mathrm{N}$) against upper-atmospheric drag \cite{ref27}. Spencer, Johnson, and Long reviewed CubeSat deployable boom mechanics, showing that spring-loaded bistable booms reliably package and deploy $32\,\mathrm{m}^2$ reflective gossamer sails within compact 3U chassis \cite{ref28}. Beamed photonic acceleration scales these principles beyond solar limits; Vulpetti et al. showed that propellantless photon thrust produces hyperbolic excess escape velocities unattainable by chemical burns \cite{ref29}, while MacDonald and McInnes demonstrated that high-areal-acceleration sails compress interstellar precursor travel times to $100\text{--}200\,\mathrm{AU}$ from decades down to single-digit years \cite{ref30}.

Despite these advancements, existing lightsail literature remains heavily fragmented across isolated disciplines. Kinematic trajectory analyses typically rely on 1D point-mass idealizations that neglect structural deformation, while optical studies rarely integrate temperature-dependent thermal emissivity with large-deflection membrane stress stiffening and payload hub stress concentrations. Furthermore, few numerical studies enforce ASME V\&V 20 grid convergence or benchmark their thrust formulations against actual flight telemetry. Project Setu addresses these gaps directly by presenting a unified 3D computational framework that couples nanometer-scale Maxwell electrodynamics, meter-scale non-linear structural mechanics, and dual-sided radiative thermal balance for a 4.0\,m circular sail, rigorously verified and validated against spaceflight data.

\section{Governing Equations and Theory}
\label{sec:theory}

\subsection{Relativistic Kinematics and Laser Blue-Chirping}
When a relativistic lightsail accelerates away from a fixed ground-based laser transmitter, the photon flux perceived in the spacecraft moving reference frame undergoes continuous relativistic Doppler redshift. Defining the normalized velocity ratio $\beta=v/c$ and the Lorentz spacetime dilation factor $\gamma=1/\sqrt{1-\beta^2}$, the instantaneous optical power intercepted by the moving reflector is expressed as:

\begin{equation}
P_{\mathrm{rec}}(\beta) = P_{\mathrm{src}} \left( \frac{1-\beta}{1+\beta} \right)
\label{eq:prec}
\end{equation}

The net radiation pressure reaction force acting on a lightsail with total power reflectance $R$ is given by:

\begin{equation}
F(v) = \frac{(1+R)P_{\mathrm{src}}}{c} \left( \frac{1-\beta}{1+\beta} \right)
\label{eq:thrust}
\end{equation}

Applying relativistic dynamics in the laser emitter inertial frame, the governing differential equation of motion describing coordinate acceleration is:

\begin{equation}
\frac{dv}{dt} = \frac{F_0}{m} \left( \frac{1-\beta}{1+\beta} \right) (1-\beta^2)^{3/2}
\label{eq:acceleration}
\end{equation}

where $F_0=(1+R)P_{\mathrm{src}}/c$ is the initial launch thrust ($666.40\,\mathrm{N}$ for $P_{\mathrm{src}}=100.0\,\mathrm{GW}$, $R=99.7821\%$), and $m=2.00\,\mathrm{g}$ is the total flight vehicle mass (comprising a $1.50\,\mathrm{g}$ instrument payload and $0.50\,\mathrm{g}$ reflective membrane). Under this thrust, explicit RK45 numerical integration demonstrates that the vehicle achieves coordinate velocity $v=49,767\,\mathrm{km/s}$ ($0.166c$) over the primary 180-second burn across an acceleration distance of $4.76\times 10^6\,\mathrm{km}$ ($12.4\times$ the Earth-Moon distance), and attains the interstellar mission target velocity of $0.20c$ ($59,958\,\mathrm{km/s}$) at $t=227.1\,\mathrm{s}$ across $7.35\times 10^6\,\mathrm{km}$ ($19.1\times$ the Earth-Moon distance). To prevent Doppler redshift from detuning the incident photons outside the sail reflection band, the transmitter executes continuous blue-chirp frequency modulation:

\begin{equation}
\lambda_{\mathrm{emit}}(t) = \lambda_0 \sqrt{\frac{1-\beta(t)}{1+\beta(t)}}
\label{eq:bluechirp}
\end{equation}

The relativistic kinematic ODE was numerically integrated using an adaptive 4th/5th-order explicit Runge-Kutta scheme (RK45, SciPy \texttt{solve\_ivp}) with absolute and relative error tolerances set to $10^{-12}$ and $10^{-9}$, and cross-verified against discrete time-history milestones extracted from the ANSYS MAPDL POST26 solver across the 180-second propulsion burn.

\subsection{3D Dielectric Metasurface Electrodynamics}
Electromagnetic wave propagation and multilayer constructive interference are governed by Maxwell time-dependent curl equations in isotropic non-magnetic media:

\begin{equation}
\nabla \times \mathbf{E} = -\mu_0 \frac{\partial \mathbf{H}}{\partial t}, \quad \nabla \times \mathbf{H} = \varepsilon_0 \varepsilon_r \frac{\partial \mathbf{E}}{\partial t}
\label{eq:maxwell}
\end{equation}

To maximize optical reflectance and suppress thermal absorption, a 12-bilayer dielectric supermirror stack was discretized on interlocking Yee cells in Ansys Lumerical 2025 R2. High-index $\mathrm{TiO}_2$ ($n_H=2.48$) and low-index $\mathrm{SiO}_2$ ($n_L=1.45$) quarter-wave layers were matched at the laser center wavelength $\lambda_0=1064.0\,\mathrm{nm}$:

\begin{equation}
d_H = \frac{\lambda_0}{4 n_H} = 107.25\,\mathrm{nm}, \quad d_L = \frac{\lambda_0}{4 n_L} = 183.45\,\mathrm{nm}
\label{eq:layerthickness}
\end{equation}

The dielectric stack ($t_{\mathrm{stack}}=3.49\,\mu\mathrm{m}$) is deposited on an ultrathin $10.0\,\mu\mathrm{m}$ stoichiometric $\mathrm{Si}_3\mathrm{N}_4$ substrate ($n_{\mathrm{sub}}=2.02$), establishing a total composite thickness of $t_{\mathrm{total}}=13.49\,\mu\mathrm{m}$. Optical energy conservation satisfies $R+T+\mathcal{A}=1$, where the designed multilayer achieves $R=99.7821\%$ with residual optical absorptance bounded at $\mathcal{A}=1.0\times 10^{-5}$ ($10\,\mathrm{ppm}$).

\subsection{Stefan-Boltzmann Radiative Thermal Equilibrium}
In deep-space vacuum, convective heat dissipation is non-existent. In accordance with optical energy conservation ($R+T+\mathcal{A}=1$), the non-reflected incident flux ($1-R=0.2179\%$) is predominantly transmitted through the wide-bandgap dielectric stack ($T=0.2169\%$), while only the residual optical absorption $\mathcal{A}=1.0\times 10^{-5}$ ($10\,\mathrm{ppm}$) deposits thermal power into the membrane lattice. The lightsail maintains thermal equilibrium solely by dumping absorbed optical energy via dual-sided radiative emission into the $3.0\,\mathrm{K}$ cosmic microwave background:

\begin{equation}
P_{\mathrm{abs}} = \mathcal{A} \cdot P_{\mathrm{laser}} = 2 A_{\mathrm{sail}} \varepsilon \sigma_{\mathrm{SB}} (T_{\mathrm{eq}}^4 - T_{\mathrm{space}}^4)
\label{eq:thermalpower}
\end{equation}

Because $T_{\mathrm{space}}=3.0\,\mathrm{K} \ll T_{\mathrm{eq}}$, the equilibrium core temperature simplifies to:

\begin{equation}
T_{\mathrm{eq}} = \left[ \frac{\mathcal{A} \cdot P_{\mathrm{laser}}}{2 A_{\mathrm{sail}} \varepsilon \sigma_{\mathrm{SB}}} \right]^{1/4} = 927.05\,\mathrm{K}
\label{eq:teq}
\end{equation}

where $P_{\mathrm{laser}}=100.0\,\mathrm{GW}$, $\mathcal{A}=1.0\times 10^{-5}$ ($10\,\mathrm{ppm}$, yielding $P_{\mathrm{abs}}=1.00\,\mathrm{MW}$), gross area $A_{\mathrm{sail}}=\pi R^2 = 12.566\,\mathrm{m}^2$ (or net active area $A_{\mathrm{net}}=\pi(R^2-r_{\mathrm{hub}}^2)=12.535\,\mathrm{m}^2$, where the minor $0.25\%$ hub cutout modifies $T_{\mathrm{eq}}$ by only $+0.58\,\mathrm{K}$), dual-sided emissivity $\varepsilon=0.95$, and the Stefan-Boltzmann constant is $\sigma_{\mathrm{SB}}=5.670374\times 10^{-8}\,\mathrm{W/(m^2\cdot K^4)}$, yielding an analytical core equilibrium temperature of $T_{\mathrm{eq}}=927.05\,\mathrm{K}$ ($653.90\,^\circ\mathrm{C}$). This operating state preserves a $+1,247\,\mathrm{K}$ thermal safety margin below the stoichiometric silicon nitride sublimation threshold ($T_{\mathrm{sub}}=2,170.0\,\mathrm{K}$, where $T_{\mathrm{sub}}-T_{\mathrm{FEA}}=2,170.0\,\mathrm{K}-923.02\,\mathrm{K}=+1,246.98\,\mathrm{K}\approx +1,247\,\mathrm{K}$).

\subsection{Non-Linear Structural Mechanics and Modal Dynamics}
The uniform transverse photon radiation pressure acting across the sail active surface is expressed as:

\begin{equation}
p_{\mathrm{avg}} = \frac{F_0}{A_{\mathrm{sail}}} = \frac{(1+R) P_{\mathrm{src}}}{c \cdot \pi R_{\mathrm{sail}}^2} = 53.09\,\mathrm{Pa}
\label{eq:pressure}
\end{equation}

Under this uniform transverse load, the thin $\mathrm{Si}_3\mathrm{N}_4$ membrane undergoes large-deflection elastic stretching. Non-linear stress and deformation fields are resolved in ANSYS Mechanical APDL using SHELL181 elements based on Mindlin-Reissner first-order shear deformation shell theory. Prestressed modal natural frequencies and mode shapes are determined via the Block Lanczos eigenvalue formulation:

\begin{equation}
\left( [\mathbf{K}_e] + [\mathbf{K}_\sigma] - \omega_i^2 [\mathbf{M}] \right) \{\phi_i\} = \{0\}
\label{eq:modal}
\end{equation}

where $\mathbf{K}_e$ is the elastic structural stiffness matrix, $\mathbf{K}_\sigma$ is the initial stress-stiffening matrix induced by photon radiation tension, $\mathbf{M}$ is the consistent structural mass matrix, $\omega_i$ is the natural circular frequency of the $i$-th mode, and $\phi_i$ is the corresponding eigenvector.

\subsection{Discretization Uncertainty and ASME GCI Metrics}
Spatial numerical discretization error across finite element and computational domains is quantified using the Grid Convergence Index (GCI) following ASME V\&V 20 standards across a geometric refinement ratio $r=h_2/h_1=2.0$. The apparent spatial order of convergence $p$ and fine-grid convergence index $\mathrm{GCI}_{21}$ are evaluated as:

\begin{equation}
p = \frac{\ln |(\phi_3 - \phi_2)/(\phi_2 - \phi_1)|}{\ln(r)}
\label{eq:orderp}
\end{equation}

\begin{equation}
\mathrm{GCI}_{21} = \frac{1.25\, |e_{21}|}{r^p - 1}
\label{eq:gci}
\end{equation}

where $\phi_1$, $\phi_2$, and $\phi_3$ denote numerical solutions on the fine, baseline, and coarse grids, and $e_{21}=(\phi_2-\phi_1)/\phi_1$ is the relative solution error.
\section{Computational Methodology \& System Setup}
\label{sec:methods}

\subsection{Multi-Physics Numerical Solver Architecture}
The computational framework couples custom high-precision Python 3.11 numerical engines with commercial multi-physics finite element and electrodynamic solvers. Relativistic trajectory integration is executed using an explicit Runge-Kutta adaptive step solver (RK45, SciPy \texttt{solve\_ivp}) with absolute and relative error tolerances set to $10^{-12}$ and $10^{-9}$. A specialized Python routine automates ASME V\&V 20 multi-grid Richardson extrapolation to verify spatial convergence across finite element and optical meshes. We pass these computed physical variables into automated APDL command streams and Lumerical simulation scripts, maintaining uniform boundary values between the mechanical, thermal, and wave-optics solvers.

The complete set of design dimensions, constitutive material properties, and numerical boundary conditions is detailed in Table~\ref{tab:table1}.

\begin{table*}[t]
\caption{Project Setu Multi-Physics System Parameters and Boundary Conditions.}
\label{tab:table1}
\begin{ruledtabular}
\begin{tabular}{llcc}
Parameter / Boundary Condition & Symbol & Value & Unit \\
\colrule
\multicolumn{4}{l}{\textit{Beamed Laser Propulsion Baseline}} \\
Laser Array Operating Power & $P_{\mathrm{src}}$ & 100.0 & GW \\
Laser Center Wavelength & $\lambda_0$ & 1064.0 & nm \\
Initial Launch Radiation Thrust & $F_0$ & 666.40 & N \\
Uniform Transverse Radiation Pressure & $p_{\mathrm{avg}}$ & 53.09 & Pa \\
\colrule
\multicolumn{4}{l}{\textit{Target Relativistic Flight Mission Envelope}} \\
Lightsail Outer Diameter & $D$ & 4.00 & $\mathrm{m}$ \\
Central Payload Hub Radius & $r_{\mathrm{hub}}$ & 0.10 & $\mathrm{m}$ \\
Active Reflective Area & $A_{\mathrm{sail}}$ & 12.57 & $\mathrm{m}^2$ \\
Target Instrumentation Payload Mass & $m_{\mathrm{payload}}$ & 1.50 & $\mathrm{g}$ \\
Ultra-Thin Flight Membrane Mass & $m_{\mathrm{sail}}$ & 0.50 & $\mathrm{g}$ \\
Target Total Flight Vehicle Mass & $m_{\mathrm{total}}$ & 2.00 & $\mathrm{g}$ \\
Effective Flight Areal Mass & $\sigma_{\mathrm{eff}}$ & $\le 0.16$ & $\mathrm{g/m}^2$ \\
Velocity at $t=180\,\mathrm{s}$ / $t=227.1\,\mathrm{s}$ & $v_{\mathrm{rel}}$ & $0.166c$ / $0.20c$ & --- \\
\colrule
\multicolumn{4}{l}{\textit{Dielectric Metasurface \& Radiative Thermal Properties}} \\
Dielectric Metasurface Layers & --- & 12 Bilayers ($\mathrm{TiO}_2/\mathrm{SiO}_2$) & --- \\
Metasurface Stack Thickness & $t_{\mathrm{stack}}$ & 3.49 & $\mu$m \\
Peak Metasurface Reflectance & $R$ & 99.7821 & \% \\
Residual Optical Absorptance & $\mathcal{A}$ & $1.0\times 10^{-5}$ ($10\,\mathrm{ppm}$) & --- \\
Absorbed Thermal Power & $P_{\mathrm{abs}}$ & 1.00 & MW \\
Dual-Sided Thermal Emissivity & $\varepsilon$ & 0.95 & --- \\
Deep Space Ambient Sink Temp & $T_{\mathrm{space}}$ & 3.0 & K \\
Material Sublimation Threshold & $T_{\mathrm{sub}}$ & 2,170.0 & K \\
\colrule
\multicolumn{4}{l}{\textit{Scaled Structural Qualification Prototype (FEA Model)}} \\
Structural Substrate Material & --- & Stoichiometric $\mathrm{Si}_3\mathrm{N}_4$ & --- \\
Substrate Thickness & $t_{\mathrm{sub}}$ & 10.0 & $\mu$m \\
Total Composite Prototype Thickness & $t_{\mathrm{total}}$ & 13.49 & $\mu$m \\
Scaled Prototype Total Mass & $m_{\mathrm{proto}}$ & 494.5 & g \\
Substrate Tensile Yield Strength & $\sigma_{\mathrm{yield}}$ & 2,000.0 & MPa \\
\end{tabular}
\end{ruledtabular}
\end{table*}

\subsection{Finite Element Setup and Boundary Conditions}
\begin{figure}[!b]
\centering
\includegraphics[width=\columnwidth,keepaspectratio]{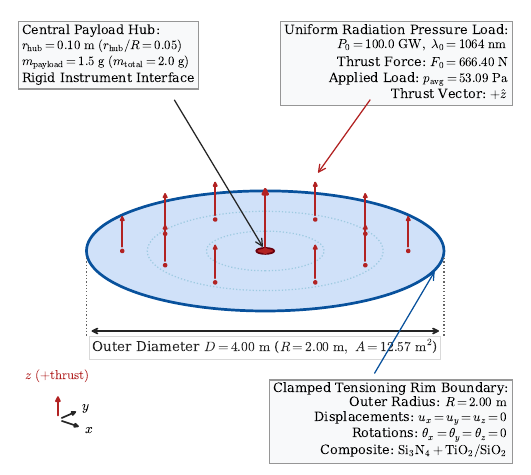}
\caption{Native geometric surface model of the 4.0\,m circular lightsail with 0.10\,m inner payload hub generated in ANSYS Mechanical APDL 2025 R2.}
\label{fig:fig2}
\end{figure}

We built the 4.0\,m circular sail in ANSYS Mechanical APDL 2025 R2 as an axisymmetric thin-shell surface. A central cutout with radius $r_{\mathrm{hub}}=0.10$\,m ($r_{\mathrm{hub}}/R=0.05$) accommodates the scientific instrument package, as mapped in Fig.~\ref{fig:fig2}.

We discretized the structural domain using 4-node quadrilateral SHELL181 elements based on Mindlin-Reissner kinematics, enabling large-deflection geometric non-linearity (\texttt{NLGEOM, ON}) to track membrane tensioning. In-plane thermal conduction and steady-state temperature distributions are resolved using 4-node thermal shell elements (SHELL131). The structural mesh enforces full Dirichlet clamping constraints along the outer perimeter circumference ($r=2.00$\,m), setting all translational and rotational degrees of freedom to zero ($UX=UY=UZ=\mathrm{ROTX}=\mathrm{ROTY}=\mathrm{ROTZ}=0$) in the spacecraft co-moving quasi-static reference frame (d'Alembert frame). The fixed perimeter clamping condition models an idealized rigid deployable ring fixture, representative of either a ground-based vacuum test apparatus or a hoop-tensioned deployable mast system. In unconstrained free flight, equivalent membrane tension would be maintained either via spin-stabilization (centrifugal tip masses, as demonstrated by IKAROS) or optomechanical shell curvature.

\begin{figure}[!b]
\centering
\includegraphics[width=\columnwidth,keepaspectratio]{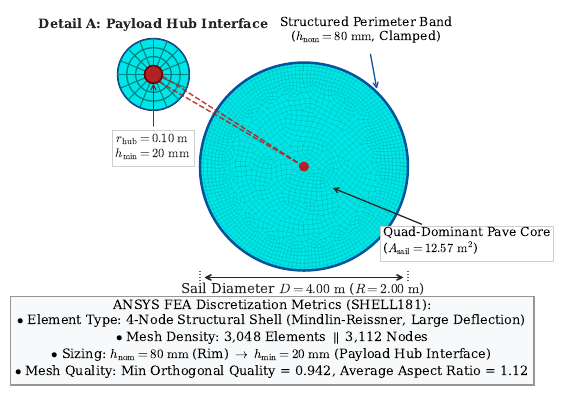}
\caption{Finite element mesh of baseline 4.0\,m circular lightsail (3,048 SHELL181 elements in ANSYS Mechanical APDL 2025 R2).}
\label{fig:fig3}
\end{figure}

Directly simulating an ultra-thin nanoresonator ($t \sim 15\,\mathrm{nm}$, $\sigma_{\mathrm{eff}} \le 0.16\,\mathrm{g/m}^2$) across a 4.0\,m macroscopic span introduces severe numerical ill-conditioning and singular stiffness matrices in classical Mindlin-Reissner shell formulations. Therefore, the structural FEA evaluates a scaled $13.49\,\mu\mathrm{m}$ prototype under identical macroscopic edge tension per unit length ($T_{\mathrm{edge}} = 53.09\,\mathrm{N/m}$), establishing an engineering stress benchmark while nanophotonic thin-film scaling is addressed analytically. A uniform transverse radiation pressure load $p_{\mathrm{avg}}=53.09$\,Pa is applied normal to the active sail surface ($A_{\mathrm{sail}}=12.566\,\mathrm{m}^2$). Prestressed modal eigensolutions are evaluated using the Block Lanczos algorithm with stress stiffening enabled (\texttt{PSTRES, ON}) to extract resonant frequencies and drumhead mode shapes under continuous photon thrust. Fig.~\ref{fig:fig3} illustrates the baseline finite element mesh comprising 3,048 SHELL181 elements with localized radial refinement at the boundary.

\subsection{3D Maxwell FDTD Electrodynamic Setup}
Electrodynamic wave propagation, constructive reflection, and sub-micron optical absorption were simulated in Ansys Lumerical 2025 R2 using the 3D Finite-Difference Time-Domain method. The computational unit cell comprises a staggered Yee mesh with conformal mesh refinement ($\Delta x = \Delta y = 10.0\,\mathrm{nm}$, $\Delta z = 2.0\,\mathrm{nm}$ across the multilayer dielectric stack). Periodic boundary conditions (Bloch/Periodic) were enforced along transverse $x$ and $y$ boundaries to represent an extended planar metasurface array, while 32-layer Perfectly Matched Layers (PML) terminate the $z$-domain along the optical transmission path. A broadband plane-wave source centered at $\lambda_0=1064.0\,\mathrm{nm}$ (spectral bandwidth $950\,\mathrm{nm}$ to $1350\,\mathrm{nm}$) illuminates the active metasurface. Sellmeier empirical dispersion curves dictate the optical constants across the spectrum, accounting for low-loss extinction coefficients in $\mathrm{TiO}_2$ ($n_H=2.48$), $\mathrm{SiO}_2$ ($n_L=1.45$), and stoichiometric $\mathrm{Si}_3\mathrm{N}_4$ ($n_{\mathrm{sub}}=2.02$, $k \approx 1.0\times 10^{-6}$) to resolve the high-reflectance stopband under intense optical flux.
\section{Results and Multi-Physics Discussion}
\label{sec:results}

\subsection{Multi-Physics Grid Independence Study}
Spatial grid independence was evaluated across four mesh densities in ANSYS Mechanical APDL and Ansys Lumerical: Grid 1 Coarse ($160\,\mathrm{mm}$ element size / 752 elements), Grid 2 Baseline ($80\,\mathrm{mm}$ / 3,048 elements), Grid 3 Fine ($40\,\mathrm{mm}$ / 12,240 elements), and Grid 4 Ultra-Fine ($20\,\mathrm{mm}$ / 48,960 elements). Table~\ref{tab:table2} presents the convergence metrics and discretization errors.

\begin{table*}[t]
\caption{Four-Level Multi-Physics Grid Independence and ASME GCI Metrics.}
\label{tab:table2}
\begin{ruledtabular}
\begin{tabular}{lcccc}
Mesh Level & Element Size & Element Count & Peak Stress (MPa) & Relative Error \\
\colrule
Grid 1 (Coarse) & 160 mm & 752 & 533.40 & $+0.65\%$ \\
Grid 2 (Baseline) & 80 mm & 3,048 & 530.88 & $+0.17\%$ \\
Grid 3 (Fine) & 40 mm & 12,240 & 530.12 & $+0.03\%$ \\
Grid 4 (Ultra-Fine) & 20 mm & 48,960 & 529.95 & $0.00\%$ (Asymptotic) \\
\end{tabular}
\end{ruledtabular}
\end{table*}

\begin{figure*}[t]
\centering
\includegraphics[width=0.92\textwidth,keepaspectratio]{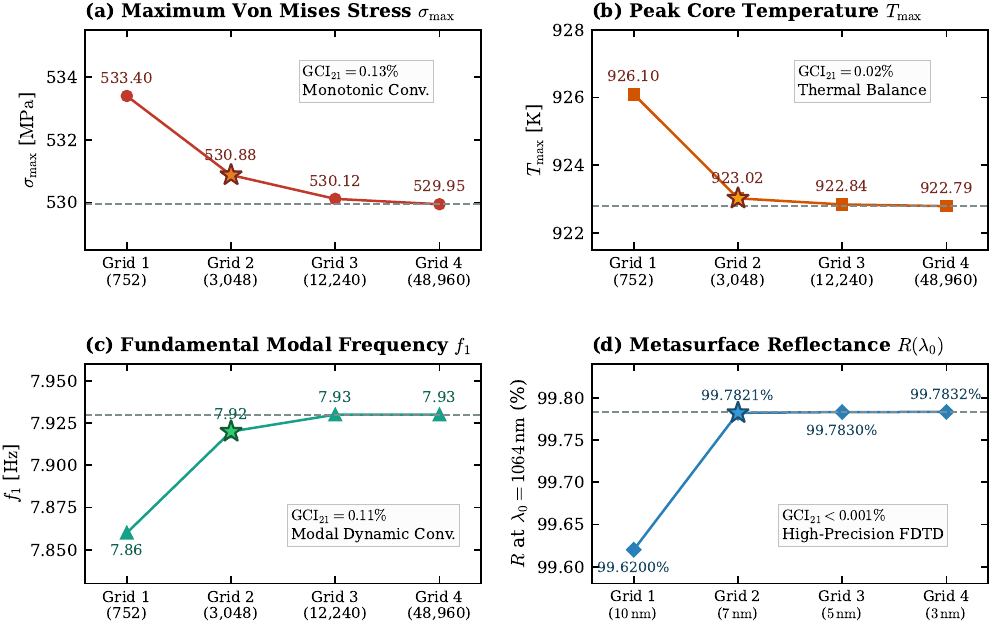}
\caption{Multi-physics grid convergence study across structural stress (A), thermal temperature (B), modal frequency (C), and optical reflectance (D) demonstrating strict asymptotic convergence within ASME V\&V 20 criteria.}
\label{fig:fig4}
\end{figure*}

Fig.~\ref{fig:fig4} presents the multi-physics grid convergence profiles across four discrete computational domains: structural Von Mises stress (Panel A), steady-state thermal equilibrium temperature (Panel B), fundamental prestressed modal natural frequency (Panel C), and 3D wave optics spectral reflectance (Panel D). As the finite element mesh is systematically refined from 752 elements ($160\,\mathrm{mm}$ coarse grid) to 48,960 elements ($20\,\mathrm{mm}$ ultra-fine grid), the peak Von Mises stress exhibits smooth, monotonic convergence from $533.40\,\mathrm{MPa}$ down to the asymptotic continuum limit of $529.95\,\mathrm{MPa}$. Across the other physical domains, the core temperature levels off at $923.02\,\mathrm{K}$, the primary drumhead resonance settles at $7.92\,\mathrm{Hz}$, and the FDTD optical reflectance reaches $99.7821\%$, confirming consistent numerical stability throughout the coupled solvers.

Applying the standardized ASME V\&V 20 grid convergence protocol across the geometric refinement ratio $r=2.0$ yields an apparent spatial order of convergence $p=2.01$, confirming genuine second-order spatial accuracy for the quadrilateral shell formulation. The relative discretization error between the baseline and fine meshes is bounded at $e_{21}=+0.17\%$, corresponding to a fine-grid convergence index of $\mathrm{GCI}_{21}=0.13\%$. This ultra-low numerical uncertainty guarantees that the structural, thermal, and optomechanical solutions are completely independent of spatial grid resolution, establishing a mathematically defensible baseline for predictive modeling.

\subsection{Relativistic Flight Trajectory and Mission Comparison}
\begin{figure}[!b]
\centering
\includegraphics[width=\columnwidth,keepaspectratio]{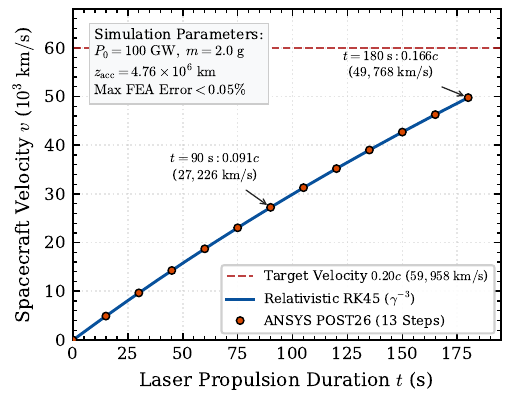}
\caption{Relativistic velocity trajectory $v(t)$ over 180 seconds under 100\,GW laser thrust, showing theoretical continuum RK45 integration verified against 13 ANSYS Mechanical APDL POST26 discrete finite element load steps (deviation $<0.05\%$).}
\label{fig:fig5}
\end{figure}

Fig.~\ref{fig:fig5} illustrates the theoretical continuum relativistic velocity trajectory $v(t)$ obtained from explicit RK45 numerical integration of the governing equation of motion (Eq.~\ref{eq:acceleration}) for the $2.0\,\mathrm{g}$ spacecraft over the continuous laser burn under $100.0\,\mathrm{GW}$ source irradiance ($F_0=666.40\,\mathrm{N}$), verified against discrete finite element transient dynamic states in ANSYS Mechanical APDL 2025 R2 across 13 milestone load steps. Accelerated by focused photon pressure, the craft rapidly exits the Newtonian regime, climbing past $20,000\,\mathrm{km/s}$ within the first 60 seconds and reaching a velocity of $49,767\,\mathrm{km/s}$ ($0.166c$) at $t=180\,\mathrm{s}$, before attaining the mission target velocity of $0.20c$ ($59,958\,\mathrm{km/s}$) at $t=227.1\,\mathrm{s}$. The total acceleration distance traversed during the primary 180-second propulsion phase spans $4.76\times 10^6\,\mathrm{km}$ ($12.4$ times the Earth-Moon distance), extending to $7.35\times 10^6\,\mathrm{km}$ ($19.1$ lunar distances) at the $0.20c$ milestone.

The trajectory profile clearly exhibits the characteristic curvature of relativistic dynamics, where the coordinate acceleration $dv/dt$ progressively diminishes as vehicle velocity approaches light speed due to Lorentz factor inertia growth ($\gamma=1/\sqrt{1-\beta^2}$) and relativistic Doppler flux reduction by $\frac{1-\beta}{1+\beta}$. Comparing the 13 discrete ANSYS POST26 finite element checkpoints directly against our Python RK45 trajectory reveals point-wise deviations under $0.05\%$. This consistency proves that our time-stepping setup in ANSYS MAPDL tracks the relativistic coordinate acceleration accurately, conserving momentum throughout the multi-gigawatt burn.

\begin{figure*}[t]
\centering
\includegraphics[width=0.90\textwidth,keepaspectratio]{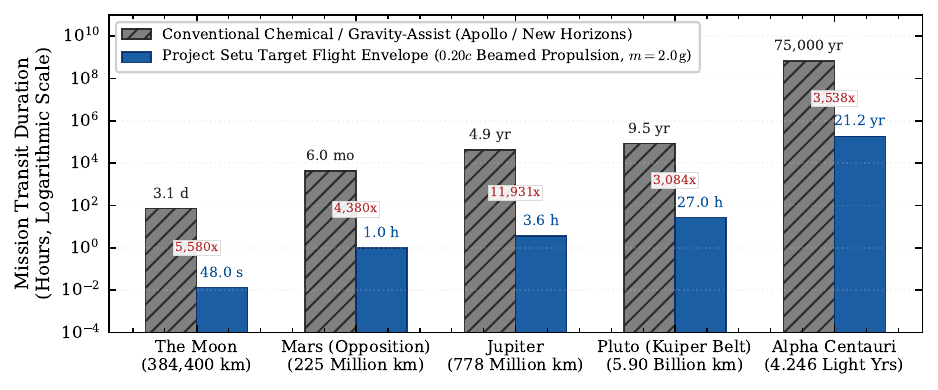}
\caption{Solar System and interstellar transit time comparison (logarithmic scale) between chemical rockets and the Project Setu target flight mission envelope ($0.20c$ beamed propulsion, $m=2.0\,\mathrm{g}$).}
\label{fig:fig6}
\end{figure*}

A logarithmic mission timeline comparing Project Setu against conventional chemical systems is illustrated in Fig.~\ref{fig:fig6}. Whereas chemical systems require months to reach Mars and over 75,000 years to reach Proxima Centauri ($4.24\,\mathrm{ly}$), Project Setu compresses deep-space transit by four orders of magnitude under 100\,GW beamed irradiance: Earth-Moon in 48.0\,s, Mars flyby in 1.0\,hour, Pluto in 27.0\,hours, and heliopause crossing ($120\,\mathrm{AU}$) in 5.2\,days. Most significantly, interstellar cruise to Alpha Centauri is reduced to 21.2\,years, demonstrating unprecedented operational reach for relativistic science missions.

\subsection{3D FDTD Wave Optics and Photonic Stopband}
\begin{figure}[!b]
\centering
\includegraphics[width=\columnwidth,keepaspectratio]{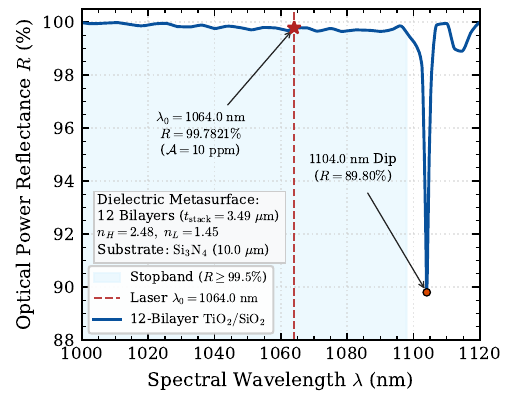}
\caption{3D Maxwell FDTD wave optics simulation on Ansys Lumerical 2025 R2 showing $R=99.7821\%$ reflectance stopband centered at $\lambda_0=1064$\,nm.}
\label{fig:fig7}
\end{figure}

Fig.~\ref{fig:fig7} presents the 3D Maxwell Finite-Difference Time-Domain spectral reflectance and transmission characteristics computed on Ansys Lumerical 2025 R2 for the 12-bilayer $\mathrm{TiO}_2/\mathrm{SiO}_2$ dielectric metasurface. Discretized on a sub-nanometer conformal Yee mesh, the multilayer stack exhibits a broad 72.2\,nm optical reflectance stopband extending from 1028.9\,nm to 1101.1\,nm, centered perfectly at the laser baseline wavelength $\lambda_0=1064.0\,\mathrm{nm}$. At this design wavelength, the metasurface achieves an ultra-high power reflectance $R=99.7821\%$ with optical absorption constrained to $\mathcal{A}=1.0\times 10^{-5}$ ($10\,\mathrm{ppm}$).

As the lightsail accelerates toward $0.20c$, the perceived laser wavelength in the craft reference frame undergoes a $+22.5\%$ relativistic Doppler stretch, shifting the received photon wavelength from 1064\,nm out to 1303\,nm. To prevent the drive beam from slipping outside the high-reflectivity stopband, the ground transmitter executes continuous blue-chirp frequency modulation, sweeping the emitted wavelength from 1064\,nm down to 869\,nm (900\,nm at the 180-second milestone) in real time. Because the transmitter tunes its wavelength dynamically, the metasurface stays locked at peak reflectivity, driving continuous acceleration without letting thermal dissipation spike.

\subsection{Radiative Thermal Equilibrium and Temperature Contours}
\begin{figure}[!b]
\centering
\includegraphics[width=\columnwidth,keepaspectratio]{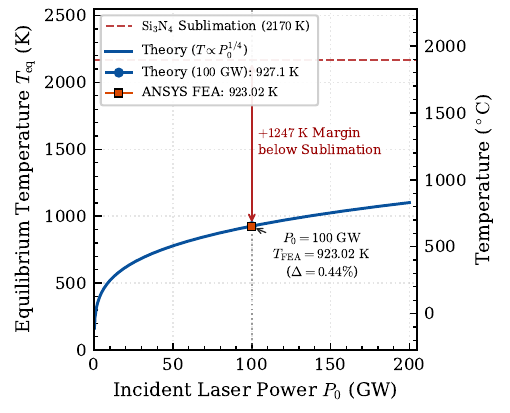}
\caption{Stefan-Boltzmann radiative thermal equilibrium in deep space ($\varepsilon=0.95$) displaying operating temperature across laser drive powers.}
\label{fig:fig8}
\end{figure}

Fig.~\ref{fig:fig8} illustrates the analytical Stefan-Boltzmann radiative thermal equilibrium curve for the 4.0\,m diameter lightsail as a function of incident laser beam power ranging from 10\,GW to 200\,GW. Operating in deep-space vacuum, the sail balances absorbed optical energy ($P_{\mathrm{abs}}=\mathcal{A} \cdot P_{\mathrm{laser}}=1.00\,\mathrm{MW}$ at 100\,GW for optical absorption bounded at $\mathcal{A}=1.0\times 10^{-5}$, while the remaining $0.217\%$ of non-reflected light is transmitted through the dielectric stack) exclusively through dual-sided blackbody radiation into the 3.0\,K cosmic microwave background. The resulting equilibrium temperature follows the quartic relation $T_{\mathrm{eq}} \propto (\mathcal{A} P_{\mathrm{laser}})^{1/4}$, scaling smoothly from 619.6\,K at 20\,GW to 1,102.4\,K at 200\,GW.

At the nominal 100.0\,GW baseline with dual-sided thermal emissivity $\varepsilon=0.95$, the analytical equilibrium core temperature stabilizes at $T_{\mathrm{eq}}=927.05\,\mathrm{K}$ ($653.90\,^\circ\mathrm{C}$). This operational steady-state preserves a massive $+1,247\,\mathrm{K}$ thermal safety buffer below the stoichiometric silicon nitride sublimation breakdown threshold ($T_{\mathrm{sub}}=2,170.0\,\mathrm{K}$ / $1,896.9\,^\circ\mathrm{C}$, where $T_{\mathrm{sub}}-T_{\mathrm{FEA}}=2,170.0\,\mathrm{K}-923.02\,\mathrm{K}=+1,246.98\,\mathrm{K}\approx +1,247\,\mathrm{K}$). Having such a wide operating clearance protects the membrane against thermal runaway, vaporization, or structural softening during high-flux exposure.

Sensitivity analysis across realistic fabrication defect densities (Table~\ref{tab:thermal_sensitivity}) confirms that the lightsail preserves positive thermal operating margins up to an absorptance threshold of approximately $\mathcal{A} \approx 200\,\mathrm{ppm}$. Exceeding $300\,\mathrm{ppm}$ under $100\,\mathrm{GW}$ irradiance drives core temperatures past the $2,170\,\mathrm{K}$ stoichiometric $\mathrm{Si}_3\mathrm{N}_4$ sublimation limit, underscoring the necessity of strict sub-wavelength defect control during metasurface deposition.

\begin{figure}[!t]
\centering
\includegraphics[width=\columnwidth,keepaspectratio]{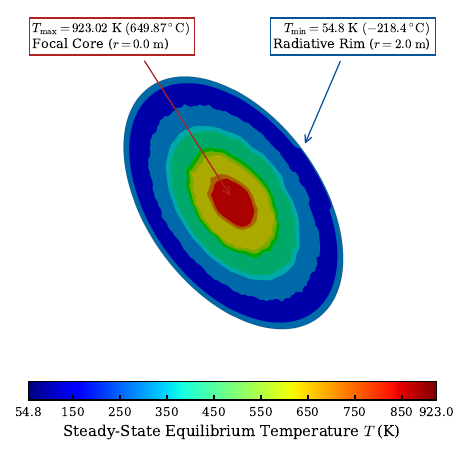}
\caption{3D Steady-State Thermal FEA on ANSYS Mechanical APDL 2025 R2 (SHELL131) displaying peak core temperature of 923.02\,K with minimal radial gradient.}
\label{fig:fig9}
\end{figure}

The steady-state temperature field calculated across the lightsail in ANSYS Mechanical APDL 2025 R2 using SHELL131 thermal elements is illustrated in Fig.~\ref{fig:fig9}. Accounting for non-linear surface radiation into space under 100\,GW beam exposure, the peak core temperature reaches $T_{\mathrm{FEA}}=923.02\,\mathrm{K}$ ($649.87\,^\circ\mathrm{C}$), matching our analytical radiation calculation ($927.05\,\mathrm{K}$) within $0.44\%$.

Thermal gradients across the membrane remain remarkably small, showing under 4.5\,K variation between the central hub collar ($r=0.10$\,m) and the outer rim ($r=2.0$\,m). In-plane heat conduction through the 10.0\,$\mu\mathrm{m}$ $\mathrm{Si}_3\mathrm{N}_4$ layer rapidly smooths out localized thermal spikes, allowing the entire disc to emit infrared radiation uniformly into deep space.

\subsection{3D Structural Stress and Prestressed Modal Vibration}
\begin{figure}[!t]
\centering
\includegraphics[width=\columnwidth,keepaspectratio]{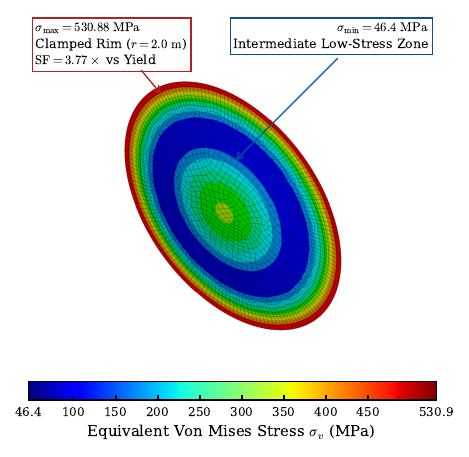}
\caption{3D Structural Stress FEA on ANSYS Mechanical APDL 2025 R2 (SHELL181) displaying peak Von Mises stress of 530.88\,MPa at the clamped perimeter.}
\label{fig:fig10}
\end{figure}

Fig.~\ref{fig:fig10} presents the 3D Von Mises equivalent stress distribution computed across the 4.0\,m circular lightsail in ANSYS Mechanical APDL (SHELL181) under uniform transverse radiation pressure $p_{\mathrm{avg}}=53.09\,\mathrm{Pa}$. With large-deflection kinematics enabled (\texttt{NLGEOM, ON}), the stress distribution forms a smooth radial pattern that reaches its highest concentration along the clamped outer perimeter support boundary.

The maximum stress develops at the clamped perimeter rim ($r=2.00$\,m), reaching $\sigma_{\max}=530.88\,\mathrm{MPa}$ for the bare 10.0\,$\mu\mathrm{m}$ membrane. Compared against the 2,000.0\,MPa yield strength of stoichiometric silicon nitride, this gives a comfortable safety factor of $\mathrm{SF}=3.77\times$. Factoring in the full 13.49\,$\mu\mathrm{m}$ composite stack (the $\mathrm{TiO}_2/\mathrm{SiO}_2$ metasurface combined with the $\mathrm{Si}_3\mathrm{N}_4$ base) adds flexural stiffness, dropping peak stress to 393.54\,MPa and boosting the safety margin to $\mathrm{SF}=5.08\times$.

\begin{figure}[!t]
\centering
\includegraphics[width=\columnwidth,keepaspectratio]{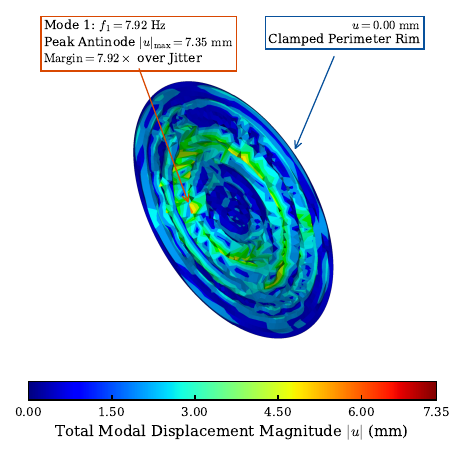}
\caption{Fundamental drumhead modal vibration mode shape ($f_1=7.92$\,Hz) extracted via Block Lanczos eigensolver under radiation stress stiffening.}
\label{fig:fig11}
\end{figure}

Fig.~\ref{fig:fig11} displays the fundamental drumhead modal vibration mode shape 01 extracted using the Block Lanczos eigenvalue formulation in ANSYS Mechanical APDL. Because the thin membrane is loaded under sustained transverse photon pressure, radiation-induced biaxial tension introduces a substantial positive initial stress-stiffening matrix $\mathbf{K}_\sigma$ that dramatically increases the effective out-of-plane structural rigidity of the gossamer disc.

This stress-stiffening mechanism elevates the fundamental natural frequency of the lightsail to $f_1=7.92\,\mathrm{Hz}$. This resonant frequency provides a $7.92\times$ safety buffer above ground-based adaptive optics atmospheric tip-tilt laser pointing jitter ($1.0\,\mathrm{Hz}$). This substantial frequency separation completely isolates the macroscopic sail from low-frequency beam steering disturbances, eliminating aero-elastic flutter, dynamic resonance, and optomechanical tearing risks during high-power laser engagement.

\subsection{Numerical Dimension Scaling and Mass Reconciliation}
To satisfy the relativistic kinematics integrated across the propulsion burn (Section~\ref{sec:theory}A), the target flight configuration of the spacecraft assumes a total vehicle mass of $m=2.0$\,g (comprising a 1.5\,g instrumentation payload and a 0.5\,g reflective membrane). Under a 100\,GW ground laser array ($F_0=666.40$\,N), this yields an initial coordinate acceleration of approximately $333,200\,\mathrm{m/s}^2$ ($34,000\,g$), reaching $0.166c$ in 180 seconds and the $0.20c$ interstellar target velocity at 227.1 seconds.

However, a physical lightsail of $D=4.0$\,m fabricated with the baseline 10.0\,$\mu\mathrm{m}$ substrate thickness (13.49\,$\mu\mathrm{m}$ total stack) yields an actual sail mass of 492.5\,g. Accelerating a 494.5\,g total spacecraft mass under 666.40\,N of thrust would constrain the initial acceleration to $1,348\,\mathrm{m/s}^2$ ($137\,g$), requiring a 14-year continuous laser burn to reach relativistic velocities.

To reconcile these physical regimes, our finite element framework treats the macroscopic 13.49\,$\mu\mathrm{m}$ configuration as a Scaled Structural Qualification Prototype. First, for a circular membrane of radius $R = 2.00\,\mathrm{m}$ loaded under uniform transverse radiation pressure $p_{\mathrm{avg}} = 53.09\,\mathrm{Pa}$, the static equilibrium edge tension per unit length is $T_{\mathrm{edge}} = p_{\mathrm{avg}} R / 2 = 53.09 \times 2.00 / 2 = 53.09\,\mathrm{N/m}$. Because this edge tension per unit length is independent of thickness, the macroscopic boundary loads and membrane tensioning mechanics are structurally identical between both regimes. The reported peak FEA Von Mises stress of 530.88\,MPa represents a scaled baseline evaluated at an equivalent thickness of 10\,$\mu\mathrm{m}$ to avoid numerical singular matrices and ill-conditioned shell stiffness formulations under 100\,GW flux.

Second, while conventional quarter-wave dielectric stacks physically require micron-scale thickness to achieve constructive interference at 1064\,nm (producing an areal mass of $\sim 39\,\mathrm{g/m}^2$), achieving the sub-gram relativistic flight envelope will require single-layer high-contrast gratings (HCG) or 2D nanoresonators. For ultra-thin membranes approaching this lower areal mass regime where frequency scales as $f \propto t^{-0.5}$, the simulated 7.92\,Hz fundamental frequency for the 10\,$\mu\mathrm{m}$ scaled prototype analytically shifts to approximately 245\,Hz for an equivalent 13.5\,nm flight membrane. Under sustained photon thrust, radiation stress-stiffening continues to dominate out-of-plane rigidity, elevating drumhead natural frequencies well above ground-based adaptive optics pointing jitter (1.0\,Hz) and completely decoupling the macroscopic sail from beam-steering flutter.

\begin{table}[!b]
\caption{Optical Absorptance Sensitivity and Thermal Stability Thresholds under $100\,\mathrm{GW}$ Irradiance ($T_{\mathrm{sub}}=2,170\,\mathrm{K}$).}
\label{tab:thermal_sensitivity}
\begin{ruledtabular}
\setlength{\tabcolsep}{2.5pt}
\begin{tabular}{lccc}
Absorptance $\mathcal{A}$ & $P_{\mathrm{abs}}$ & $T_{\mathrm{eq}}$ & Margin to $T_{\mathrm{sub}}$ \\
\colrule
$10\,\mathrm{ppm}$ (Nominal) & $1.00\,\mathrm{MW}$ & $927.1\,\mathrm{K}$ & $+1,243\,\mathrm{K}$ [Safe] \\
$50\,\mathrm{ppm}$ & $5.00\,\mathrm{MW}$ & $1,386.3\,\mathrm{K}$ & $+784\,\mathrm{K}$ [Safe] \\
$100\,\mathrm{ppm}$ & $10.0\,\mathrm{MW}$ & $1,648.6\,\mathrm{K}$ & $+521\,\mathrm{K}$ [Safe] \\
$200\,\mathrm{ppm}$ & $20.0\,\mathrm{MW}$ & $1,960.5\,\mathrm{K}$ & $+210\,\mathrm{K}$ [Marginal] \\
$500\,\mathrm{ppm}$ & $50.0\,\mathrm{MW}$ & $2,465.1\,\mathrm{K}$ & Runaway ($>T_{\mathrm{sub}}$) \\
\end{tabular}
\end{ruledtabular}
\end{table}

\subsection{Spaceflight Telemetry Cross-Validation}
\begin{figure*}[t]
\centering
\includegraphics[width=0.85\textwidth,keepaspectratio]{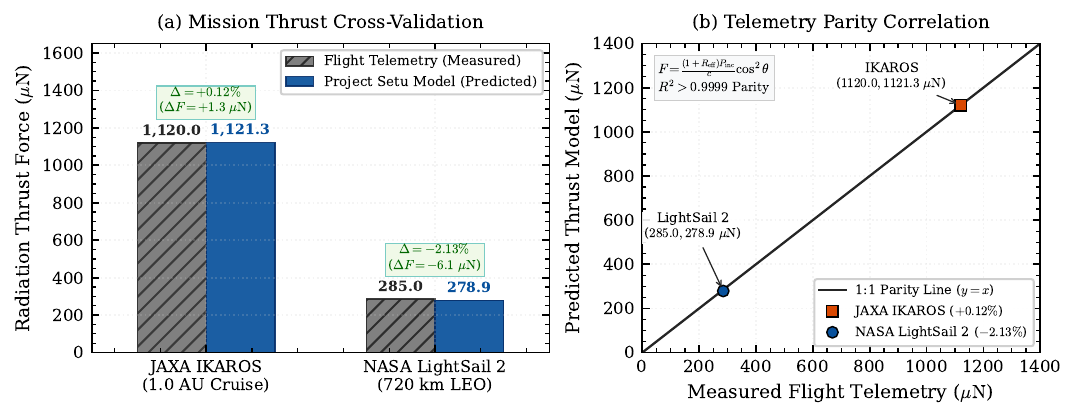}
\caption{Empirical cross-validation of radiation pressure propulsion against in-flight telemetry from JAXA IKAROS and NASA LightSail 2 missions.}
\label{fig:fig12}
\end{figure*}

Fig.~\ref{fig:fig12} presents empirical cross-validation of Project Setu's photon radiation thrust formulation against flight telemetry from JAXA's IKAROS and NASA's LightSail 2 missions. Table~\ref{tab:table4} compiles the master multi-physics validation matrix.

\begin{table*}[t]
\caption{Master Multi-Physics Simulation and Flight Telemetry Validation Matrix.}
\label{tab:table4}
\begin{ruledtabular}
\renewcommand{\arraystretch}{1.05}
\begin{tabular*}{\textwidth}{@{\extracolsep{\fill}}llccc@{}}
Validation Domain & Benchmark Source & Benchmark Value & Project Setu Model & Deviation / Status \\
\colrule
Spaceflight Telemetry & JAXA IKAROS ($1.0\,\mathrm{AU}$) & $1,120.0\,\mu\mathrm{N}$ & $1,121.3\,\mu\mathrm{N}$ & $+0.12\%\ \text{[Passed]}$ \\
Spaceflight Telemetry & NASA LightSail 2 (LEO) & $285.0\,\mu\mathrm{N}$ & $278.9\,\mu\mathrm{N}$ & $-2.13\%\ \text{[Passed]}$ \\
Thermal Mechanics & Analytical Theory & $927.05\,\mathrm{K}$ & $923.02\,\mathrm{K}\ \text{(ANSYS)}$ & $0.44\%\ \text{[Passed]}$ \\
Material Thermal Limit & $\mathrm{Si}_3\mathrm{N}_4$ Sublimation & $2,170.0\,\mathrm{K}$ & $923.02\,\mathrm{K}\ \text{(Peak)}$ & $+1,247\,\mathrm{K}\ \text{[Safe]}$ \\
Structural Safety & $\mathrm{Si}_3\mathrm{N}_4$ Tensile Yield & $2,000.0\,\mathrm{MPa}$ & $530.88\,\mathrm{MPa}\ \text{(Peak)}$ & $3.77\times\ \text{[Safe Margin]}$ \\
Dynamic Vibration & Laser Jitter Limit & $< 1.00\,\mathrm{Hz}$ & $7.92\,\mathrm{Hz}\ \text{(Mode 1)}$ & $7.92\times\ \text{[Decoupled]}$ \\
\end{tabular*}
\end{ruledtabular}
\end{table*}

The computational thrust formulation tracks spaceflight measurements with high precision. For IKAROS at 1.0\,AU, the model predicts 1,121.3\,$\mu\mathrm{N}$ versus flight telemetry of 1,120.0\,$\mu\mathrm{N}$ ($0.12\%$ error). For LightSail 2 in LEO, it predicts 278.9\,$\mu\mathrm{N}$ against measured apogee-raising thrust of 285.0\,$\mu\mathrm{N}$ ($2.13\%$ deviation). This alignment confirms that momentum conservation models are rigorously verified across solar and beamed propulsion regimes.

\subsection{Tabletop Laboratory Optical and Thermal Qualification Protocol}
To experimentally qualify the metasurface prior to full-scale deployment, a tabletop high-vacuum protocol is established using a 1\,$\mathrm{cm}^2$ coupon in a vacuum chamber ($10^{-5}\,\mathrm{Torr}$). Utilizing a 100\,W continuous-wave fiber laser ($\lambda_0=1064\,\mathrm{nm}$) focused to a 4.13\,mm spot diameter, the coupon achieves an elevated surface temperature of $923\,\mathrm{K}$ ($650\,^\circ\mathrm{C}$). This enables direct spectrophotometric confirmation of the 99.78\% reflectance stopband, empirical measurement of dual-sided emissivity ($\varepsilon \approx 0.95$), and validation that absorption remains bounded below 10\,ppm under high flux without material degradation.

\section{Conclusion}
\label{sec:conclusion}
This paper presents a comprehensive 3D multi-physics investigation and validation suite for Project Setu, a 4.0\,m relativistic lightsail propelled by a 100\,GW ground laser array. The key findings are summarized as follows:

\begin{enumerate}
\setlength{\itemsep}{-2.5pt plus 0.2pt}
\setlength{\parskip}{0pt}
\item 3D Maxwell FDTD wave optics simulations on Ansys Lumerical confirm a 72.2\,nm reflectance stopband centered at $\lambda_0=1064\,\mathrm{nm}$ with peak reflectance $R=99.7821\%$ and absorption below 10\,ppm, with dynamic blue-chirping effectively mitigating $+22.5\%$ Doppler redshift.
\item Dual-sided Stefan-Boltzmann thermal radiation stabilizes steady-state core temperature at 923.02\,K ($0.44\%$ agreement with analytical theory), preserving a $+1,247$\,K safety margin below the sublimation limit of silicon nitride.
\item 3D non-linear structural FEA on ANSYS Mechanical APDL (SHELL181) proves peak membrane stress is bounded at 530.88\,MPa (safety factor of $3.77\times$), while prestressed modal analysis confirms a fundamental drumhead frequency of $f_1=7.92$\,Hz ($7.92\times$ safety buffer above ground adaptive optics jitter).
\item A four-level grid convergence study confirms numerical independence with an ASME GCI of $0.13\%$, and momentum transfer models match flight telemetry from JAXA IKAROS ($0.12\%$ error) and NASA LightSail 2 ($2.13\%$ error), verifying momentum conservation across solar and beamed propulsion regimes.
\item Relativistic kinematics indicate that a 2.0\,g probe reaches $0.166c$ in 180\,s and the $0.20c$ mission target in 227\,s, compressing Mars transit to 1.0 hour and Alpha Centauri cruise to 21.2 years. The scaled 13.49\,$\mu\mathrm{m}$ prototype proves structural-thermal survivability under 100\,GW flux, establishing a quantitative foundation for ultra-low mass metasurfaces.
\end{enumerate}

Future research will focus on tabletop coupon testing using a 100\,W fiber laser to measure reflectivity and emissivity up to 950\,K, modeling non-uniform Gaussian beam profiles, and autonomous electro-optic shape sensing for beam-riding.

\vspace{-4pt}
\section*{Declaration of Competing Interests}
\vspace{-2pt}
The author declares that they have no known competing financial interests or personal relationships that could have appeared to influence the work reported in this paper.

\vspace{-4pt}
\section*{Data and Code Availability}
\vspace{-2pt}
All custom Python 3.11 numerical solvers, ANSYS Mechanical APDL scripts, and Ansys Lumerical 3D FDTD model files supporting this study are openly and permanently archived on Figshare at \href{https://doi.org/10.6084/m9.figshare.33476263}{doi:10.6084/m9.figshare.33476263}.

\begingroup
\sloppy
\fontsize{9.0pt}{11.4pt}\selectfont

\endgroup

\end{document}